\documentstyle[12pt]{article}
\input epsf.sty
\def\ZZZ{{\hbox{ Z\kern-1.6mm Z}}}

\newcommand{\ra}{\rangle}

\newcommand{\FF}{{\cal F}}

\newcommand{\CC}{{\cal C}}

\newcommand{\OO}{{\cal O}}

\newcommand{\wt}{\widetilde}
\newcommand{\wh}{\widehat}

\newcommand{\be}{\begin{equation}}
\newcommand{\ee}{\end{equation}}
\newcommand{\ben}{\begin{eqnarray}\displaystyle}
\newcommand{\een}{\end{eqnarray}}
\newcommand{\refb}[1]{(\ref{#1})}
\newcommand{\p}{\partial}
\newcommand{\sectiono}[1]{\section{#1}\setcounter{equation}{0}}

\def\one{{\hbox{ 1\kern-.8mm l}}}
\def\zero{{\hbox{ 0\kern-1.5mm 0}}}
 
\begin{document}
{}~
{}~
\hfill\vbox{\hbox{hep-th/0502126}
}\break
 
\vskip .6cm
\begin{center}
{\Large \bf 
Black Holes, Elementary Strings and Holomorphic Anomaly
}

\end{center}

\vskip .6cm
\medskip

\vspace*{4.0ex}
 
\centerline{\large \rm
Ashoke Sen}
 
\vspace*{4.0ex}

\centerline{\large \it Harish-Chandra Research Institute}

\centerline{\large \it  Chhatnag Road, Jhusi,
Allahabad 211019, INDIA}
 
\centerline{E-mail: ashoke.sen@cern.ch,
sen@mri.ernet.in}
 
\vspace*{5.0ex}
 
\centerline{\bf Abstract} \bigskip

In a previous paper we had proposed a specific route to relating
the entropy
of two charge black holes to the degeneracy of elementary string states
in N=4 supersymmetric heterotic
string theory in four dimensions. For toroidal compactification
this proposal works
correctly to all orders in a power series expansion in inverse
charges provided we take
into account the 
corrections to the black hole entropy formula due to holomorphic anomaly.
In this paper we demonstrate that similar agreement holds also for other
N=4 supersymmetric heterotic string compactifications.

\vfill \eject
 
\baselineskip=18pt

\tableofcontents

\sectiono{Introduction} \label{sintro}

The attempt to relate the Bekenstein-Hawking entropy of a black hole
to the number of states of the black hole in some microscopic description
of the theory is quite old. In string theory this takes a new direction
as the theory already has a large number of massive states in the spectrum
of elementary string, and hence one is tempted to speculate that for large
mass we should be able to relate the degeneracy of these states to the
entropy of a black hole with the same charge quantum 
numbers\cite{thooft,9309145,9401070,9405117,9612146}. However this
problem is complicated due to large renormalization effects which make
it difficult to relate the quantum numbers of the black hole to those
of the elementary string states. This problem can be avoided by 
considering a tower of BPS states\cite{rdabh0,rdabh1,rdabh2}
where such renormalization effects
are absent. The entropy of the corresponding
black hole solution vanishes in the supergravity approximation;
however one finds
that the curvature associated with the solution becomes large near
the horizon, indicating  breakdown of the supergravity approximation.
Although the complete analysis of the problem has not been possible to this
date, a general argument based on symmetries of the theory shows that
the entropy of the black hole, modified by $\alpha'$ corrections, has the
right dependence on all the parameters (charges as well as the asymptotic
vacuum expectation values of various fields) so as to agree with the
logarithm of the degeneracy of elementary string 
states\cite{9504147,9506200,9712150}. However the overall 
normalization constant is not determined
by the symmetry argument, and its
computation requires inclusion of all order
$\alpha'$ corrections to the tree
level heterotic string action.

It was later realized that instead of elementary strings, 
D-branes\cite{9602052} provide
a much richer arena for the study of black holes. In particular, by
considering a sufficiently complicated configuration of D-branes one can
ensure that the corresponding BPS black hole solution carrying the same 
charge
quantum numbers as the D-brane system has a finite area event
horizon where $\alpha'$ and string loop corrections are small.
Comparison of the entropy of the black hole
to the logarithm of the degeneracy of states of the D-brane configuration
(which we shall call the statistical entropy)
shows exact agreement\cite{9601029} for large charges. This
agreement has been verified for a variety of 
black holes in different string
theories.

Initial comparison between the black hole entropy and the statistical
entropy was carried out in the limit of large charges. For a class of black
holes in $N=2$
supersymmetric string compactification\cite{9508072,9602111,9602136}
ref.\cite{9711053} attempted to go beyond the large charge limit, and computed
corrections to the statistical entropy which are suppressed by the inverse
power of the charges. The corresponding corrections to the black hole entropy
come from higher derivative terms in the effective
action. By taking
into account a special class of higher derivative 
terms\cite{9602060,9603191} which come from
supersymmetrization of the curvature squared terms in the
action\cite{rzwiebach,9610237}, 
refs.\cite{9812082,9904005,9906094,9910179,0007195,0009234,0012232}
computed corrections
to the black hole entropy and found precise agreement. One non-trivial
aspect of this calculation is that in order to reach this agreement
we need to also modify the Bekenstein-Hawking formula for the black
hole entropy due to the presence of higher derivative
terms\cite{9307038,9312023,9403028,9502009}.

Recently there has been renewed interest in the black hole solution
representing elementary string states. This followed
the observation by Dabholkar\cite{0409148}
that if we take into account the special class of higher derivative
terms which were used in the analysis of 
\cite{9812082,9904005,9906094,9910179,0007195,0009234,0012232}
and calculate their
effect on the black hole solutions representing elementary string states,
we get a solution with a finite area event horizon.
The entropy of this black hole, calculated using the formul\ae\
given in \cite{9307038,9312023,9403028,9502009}, reproduces
precisely the leading term in the expression for the
statistical entropy obtained by taking the logarithm of the degeneracy
of elementary string states. This analysis was developed further in
\cite{0410076,0411255,0411272,0501014}. An alternative viewpoint for these 
black 
holes 
can be found in \cite{0412133,0502050}.

One of the advantages of using elementary string states for comparison
with black hole entropy is that for this system the degeneracy of states
and hence the statistical entropy is known very precisely. Hence one
can try to push this comparison beyond the large charge approximation.
However one problem that one encounters in this process is that even if
we know the degeneracies exactly, the
definition of the statistical entropy is somewhat ambiguous since it depends
on the particular ensemble we use to define entropy. As in the case of
an ordinary thermodynamic
system, all definitions of entropy agree when the charge (analog of volume)
is large, but finite `size' corrections to the entropy differ between the
entropies defined through different ensemble. This is due to the fact that
the agreement between different ensembles ({\it e.g.} microcanonical,
canonical and grand canonical ensembles)
is proved using a saddle point
approximation which is valid only in the `large volume' limit. Thus the
question that we need to address is: which definition of 
statistical entropy should
we use for comparison with the black hole entropy? There is no {\it
a priori} answer to this question, and one has to proceed by trial and
error to discover if there is some natural definition of
statistical entropy which agrees with the
black hole entropy beyond leading order. For a class of black holes
in $N=2$ supersymmetric string compactification,
Refs.\cite{0405146,0412139} proposed
such a definition based on a mixed ensemble where we sum over
half of the charges (the `electric' charges) by introducing
chemical potentials for these charges and keep the other half of
the charges fixed. By applying the same prescription to the black
holes representing elementary string states in $N=4$ supersymmetric
theories, \cite{0409148} was able to reproduce the black hole entropy
to all orders in the inverse charges up to an additive
logarithmic piece which
appears as a multiplicative factor in the partition function involving
powers of the winding number charge\cite{ism04}. One
disadvantage of this prescription is that it destroys manifest
symmetry between the momentum and winding charges of the string
since in defining the ensemble
we sum over all momentum states but keep fixed the winding
charge. As a result
T-duality invariance of the statistical entropy defined
this way is not guaranteed.

A related but somewhat different proposal for relating the degeneracy
of elementary string states to black hole entropy,
which maintains manifest T-duality invariance,
was proposed in \cite{0411255}. This also requires summing over
charges but in a manner that preserves manifest T-duality. In particular
the chemical potential couples to a T-duality invariant combination
of the charges.\footnote{We also sum over all angular momentum states
which is equivalent to choosing an ensemble with a chemical potential
coupled to the angular momentum, and then extremizing 
the corresponding
free energy with respect to this chemical potential. This
sets the chemical potential
to zero. This argument is due to B.~Pioline, and
I wish to thank A.~Dabholkar for discussion on this
point.}
It was shown in \cite{0411255} that up to terms which
are non-perturbative in the inverse charges, this definition
of the statistical entropy agrees with the black hole entropy including
logarithmic terms, provided we take into account the effect of
holomorphic anomaly\cite{9302103,9309140} in the effective action.
A related duality invariant
presecription for dealing with 1/4 BPS black holes in N=4
supersymmetric heterotic string compactification was later given
in \cite{0412287}. 

In order to put this proposal on a firm footing it is important that 
we test it for other N=4 heterotic string compactifications. This is
what we attempt to do in this paper. In particular we focus on a class
of four dimensional CHL models with $N=4$ 
supersymmetry\cite{CHL,CP,9507027,9507050,9508144,9508154} and compare the
statistical entropy computed using the prescription of \cite{0411255}
with the black hole entropy. We again find that after taking into account
corrections due to holomorphic anomaly, the black hole entropy and the
statistical entropy agree up to non-perturbative terms.\footnote{In
the analysis of this paper as well as in the analysis of \cite{0411255}
an overall charge independent additive constant in the expression
for the entropy could not be fixed
due to our lack of precise knowledge of the effect of the
holomorphic anomaly terms on black hole entropy.
Thus we could not compare this overall additive
constant between the black hole
and the statistical entropy.} 

The rest of the paper is organised as follows. In section \ref{srev} we
review the proposal of \cite{0411255} for relating the black hole
entropy to the degeneracy of elementary string states. In section
\ref{sstat} we review CHL string compactifications, count the degeneracy
of elementary string states in these models, and compute the statistical
entropy using these results. In section \ref{sbh} we calculate the entropy
of the black holes of the CHL model carrying the same charge quantum numbers
as the elementary string states, and show that the result agrees with the
statistical entropy found in section \ref{sstat}. During this computation
we also determine the coefficient of the holomorphic anomaly term in
these CHL models. Section \ref{sdiss} contains a discussion of
the results and possible extension to more general class of models
and/or states.
The two appendices are devoted to 
the analysis of the errors
involved in the various approximations used in this paper, and to
demonstrate that these corrections are all non-perturbative in the
inverse charges. Of the two appendices, appendix \ref{sa} analyses
the possible errors in the computation of the statistical entropy and
appendix \ref{sb} examines possible errors in the computation of
the black hole entropy. In appendix \ref{sb} we also determine the
S-duality invariant form of the curvature squared terms in the
CHL models.

I have been informed by A.~Dabholkar that for a general class of models,
ref.\cite{private}
has successfully carried out the comparison between the black
hole entropy and the entropy defined through the ensemble introduced
in \cite{0405146}. After completing this paper we also learned of
ref.\cite{9708062}
where some of the computations of section \ref{sbh} and
appendix \ref{sb}, required for determining the form of the curvature
squared terms in the effective action, have been carried out.

\sectiono{Proposal for Relating Black Hole Entropy to the Degeneracy of 
Elementary String States} \label{srev}

We shall be considering $N=4$ supersymmetric
heterotic string theory in four
dimension, with a compactification manifold of the form $K_5\times S^1$.
In this theory we consider a fundamental
string wound $w$ times along the circle $S^1$ and carrying $n$ units
of momentum along the same circle.
Let $d(n,w)$ denote the degeneracy
of elementary string states satisfying the following properties:
\begin{itemize}
\item The state is invariant under half of the space-time supersymmetry
transformations.
\item The state carries gauge charges appropriate to an elementary
heterotic string carrying $w$
units of winding and $n$ units of momentum along $S^1$. This means
that if $x^4$ denotes the coordinate along $S^1$ and $x^\mu$ ($0\le\mu\le 
3$) denote the coordinates of the non-compact part of the space-time, then 
the state carries
gauge charges proportional to $n$ and $w$ associated with the gauge fields
$G^{(10)}_{4\mu}$ and $B^{(10)}_{4\mu}$ respectively, but does 
not carry any other 
gauge charge. Here $G^{(10)}_{MN}$ and $B^{(10)}_{MN}$ denote the
ten dimensional string metric and the anti-symmetric tensor field
respectively.
\end{itemize}
We shall see in section \ref{sstat} that the
degeneracy of such states
is a function of the combination $N\equiv nw$. 
Denoting this
by $d_{N}$, we define the partition function associated with these
states as:
\be \label{e2a}
e^{\FF(\mu)} = \sum_{N} d_N \, e^{-\mu N}\, .
\ee
Given $\FF(\mu)$, we define the statistical entropy $\wt S_{stat}$ as the
Legendre transform of $\FF(\mu)$:
\be \label{e11}
\wt S_{stat}(N) = \FF(\mu) + \mu \, N\, ,
\ee
where $\mu$ is given by the solution of the equation
\be \label{e12}
{\p\FF\over \p\mu} + N =0\, .
\ee
The proposal of ref.\cite{0411255} is to identify 
$\wt S_{stat}(nw)$ with
the entropy of the black hole solution carrying same charge quantum numbers
$(n,w)$:
\be \label{eprop}
\wt S_{stat}(nw) = S_{BH}(n,w)\, .
\ee
This is the relation we shall try to verify in this paper for different
heterotic string compactifications.

The definition of statistical entropy given above is appropriate for a kind
of grand canonical ensemble where we introduce a chemical potential conjugate
to $nw$.
A 
more direct definition of the statistical entropy would be the one based on
the microcanonical ensemble:
\be \label{e12a}
S_{stat}(N) = \ln d_{N}\, .
\ee
The two definitions agree in the limit of large $N$ 
where we can evaluate
the sum in \refb{e2a} by a saddle point method. In this case the leading
contribution to $e^{\FF(\mu)}$ is given by $d_{N_0} e^{-\mu N_0}$ where
$N_0$ is the value of $N$ that maximizes the summand:
\be \label{esaddle}
\FF(\mu) \simeq \ln d_{N_0} - \mu N_0, \qquad {\p\over \p N_0}
\ln d_{N_0}-\mu = 0\, .
\ee
Thus in this approximation $\FF(\mu)$ is the Legendre transform of
$\ln d_{N_0}=S_{stat}(N_0)$. Hence $\wt S_{stat}(N)$, 
defined as the Legendre
transform of $\FF(\mu)$, will be equal to $S_{stat}(N)$.
However the complete $\FF(\mu)$ defined through \refb{e2a} has
additional contribution besides that given in \refb{esaddle},
and as a result
$S_{stat}$ and
$\wt S_{stat}$ differ in non-leading terms. In particular
the coefficient of
the $\ln N$ terms in $S_{stat}$ and 
$\wt S_{stat}$ differ. It is not
{\it a priori} clear which definition of 
statistical entropy we should be comparing
with the entropy of the black hole solution 
carrying the same quantum numbers.
It was shown in \cite{0411255} that for 
heterotic string theory compactified on a
torus, $\wt S_{stat}$ agrees with the black hole entropy
up to exponentially suppressed contributions. 
We shall see in section \ref{sbh}
that such agreement between $\wt S_{stat}$ and 
$S_{BH}$ continues to hold also
for CHL compactification\cite{CHL,CP,9508144,9508154} 
of the heterotic string theory.

Note that given $S_{stat}=\ln {d_N}$ we can calculate $\wt S_{stat}$
using eqs.\refb{e2a}-\refb{e12}. Conversely, given $\wt S_{stat}$ we
can compute $\FF(\mu)$ by taking its Legendre transform and then
compute $d_N$ by solving \refb{e2a}.
This gives $S_{stat}$. Thus $S_{stat}$
and $\wt S_{stat}$ contain complete information about each other and the 
degeneracies
$d_N$. This allows us to restate the proposal \refb{eprop} in a slightly
different but equivalent form. Given $S_{BH}(n,w)$ (which turns out to be
a function of the combination $N=nw$) we define $\FF_{BH}(\mu)$ by taking
the Legendre transform of $S_{BH}$ with respect to the variable $N$, and
then define $d^{BH}_N$ through an analog of eq.\refb{e2a} with $\FF(\mu)$ 
and 
$d_N$ replaced by $\FF_{BH}(\mu)$ and $d_N^{BH}$ respectively. The
proposal \refb{eprop} then translates to the relation:
\be \label{enewp}
\FF_{BH}(\mu) = \FF(\mu)\, , \qquad d^{BH}_N = d_N\, .
\ee
Although we shall work with \refb{eprop} for convenience, we should
keep in mind that verifying \refb{eprop} amounts to verifying
\refb{enewp}.

\sectiono{Counting Degeneracy of BPS String States in CHL Models} 
\label{sstat}

In this section we shall compute $d_N$ and hence $\FF(\mu)$ for a
class of $N=4$ supersymmetric heterotic string compactification.
First we shall illustrate the counting procedure in the context of a
specific CHL model\cite{CP} 
and then generalize this to other models. The 
construction of the model
begins with
$E_8\times E_8$ heterotic string theory compactified on a six torus
$T^4\times \wt
S^1\times S^1$. 
In this theory the gauge fields in the Cartan subalgebra consist of
22 gauge fields arising out of the left-moving U(1) currents of the
world-sheet theory, and six gauge fields arising out of the right-moving
U(1) currents of the world-sheet theory. All the gauge fields associated
with the $E_8\times E_8$ group arise out of the left-moving world-sheet
currents.
We now mod out this theory by a $Z_2$ transformation
that involves a half shift along $\wt S^1$ together with an exchange of
the two $E_8$ lattices\cite{CP}.  The resulting theory still has $N=4$
supersymmetry. In particular the 6 U(1) 
gauge fields associated with the right-moving world-sheet currents
are untouched by the $Z_2$ projection, and continue to provide
us with the graviphoton fields of $N=4$ supergravity.
On the other hand only the diagonal sum of the two $E_8$
gauge groups survive the projection. As a result the $E_8\times E_8$
component of the gauge group is reduced to $E_8$, and the rank of the
unbroken gauge group from the left-moving sector of the world-sheet is
reduced to $14$ from its original value 22. 8 of these U(1) gauge fields
come from the
surviving $E_8$ gauge group
and the other 6 come from appropriate linear combination
of the metric and antisymmetric tensor field, with one index lying
along one of the six directions of
the internal torus and the other index lying along one of the
non-compact directions.

We now consider an elementary string state wound $w$ times along $S^1$ and 
carrying $n$ units of momentum along the same $S^1$. The BPS excitations 
of this string state come from restricting the right-moving 
oscillators 
to have total level 1/2 (in the Neveu-Schwarz sector) or 0 (in the Ramond 
sector) but allowing arbitrary oscillator and momentum excitations in the 
left-moving sector. We would like to count BPS states with a given set of 
gauge charges, notably those carried by an elementary string state with 
$w$ units of winding and $n$ units of momentum along $S^1$. First let us 
do this calculation for heterotic 
string theory on a torus\cite{rdabh0}. In this case
the only possible 
excitations are those created by left-moving oscillators, since any
additional momentum and / or winding will generate additional gauge
charges carried by the state.
If $N_L$ denotes 
the total level of the left-moving oscillators then the level matching 
condition gives $N_L=nw+1$, and hence the degeneracy of elementary string 
states carrying quantum numbers $(n,w)$ is the number 
of ways we can get total oscillator level $N_L$ from the 24 left-moving 
oscillators, multiplied by a factor of 16 that counts the degeneracy of
the ground state of the right-moving sector. 
We shall call this number $d^{(0)}_{N_L}$.
It is 
given by the generating function\cite{rdabh0}
\be \label{e4}
\sum_{N_L} d^{(0)}_{N_L} e^{-\mu (N_L-1)} = 
16 \left(\eta(e^{-\mu})\right)^{-24}, \qquad
\eta(q) = q^{1/24} 
\, \prod_{n=1}^\infty ( 1 - 
q^n)\, .
\ee
For the CHL string theory under consideration, the counting is a 
little more complicated. Since only the diagonal $E_8$ gauge group 
survives, we can 
satisfy the condition for vanishing $E_8$ charge if we choose equal and 
opposite momentum vector $\vec p$ and $-\vec p$
from the two $E_8$ lattices. We choose
the overall normalization of $\vec p$ such that the $E_8$
lattice is self-dual 
under the inner product $(\vec p,\vec q)=\vec p\cdot \vec q$. In this 
normalization there is one lattice point per unit volume
in the $\vec p$ space, and the
contribution to the $\bar L_0$ eigenvalue from the vector 
$\vec s =(\vec p, -\vec p)$ is given by 
$\vec p^2/2+\vec p^2/2=\vec p^2$.
The level matching condition now gives:
\be \label{e1}
N_L + {\vec p^2} = nw + 1\, .
\ee
Thus the degeneracy $d_{nw}$ for given $(n,w)$ is equal to the number of 
ways we can satisfy \refb{e1}, subject to the condition that the
resulting state is even under the orbifold 
group:
\be \label{ednw}
d_{nw} \simeq {1\over 2} \sum_{N_L} \sum_{\vec p\in \Lambda_{E_8}} 
d^{(0)}_{N_L} \, \delta_{N_L+\vec p^2 -nw, 1}\, .
\ee
Since we only include
states which are even under the $Z_2$ transformation, we must symmetrize
the state under the exchange of the oscillators and momenta associated with
the two $E_8$ factors. As shown in appendix \ref{sa}, up to exponentially 
small contribution
this introduces a multiplicative factor of 1/2 in the counting of states
which we have included in the right hand side of \refb{ednw}.
Note that the twisted sector states do not play any
role in this counting, since they carry half-integral winding number along
the circle $\wt S^1$ and hence belong to a different charge sector. 
Using \refb{ednw} and \refb{e4} the partition function $e^{\FF(\mu)}$ 
defined in \refb{e2a} is now
given by
\be \label{e3}
e^{\FF(\mu)} \simeq {1\over 2}
\sum_{N_L} d^{(0)}_{N_L} e^{-\mu (N_L-1)} \, 
\sum_{\vec p\in \Lambda_{E_8}} 
e^{-\mu \vec 
p^2} = 8\,
\left(\eta(e^{-\mu})\right)^{-24} \, \sum_{\vec p\in \Lambda_{E_8}} 
e^{-\mu \vec 
p^2}\, .
\ee

We shall be interested in the behaviour of $\FF(\mu)$ at small $\mu$. In this
limit,
\be \label{e6}
\left(\eta(e^{-\mu})\right)^{-24} \simeq e^{{4\pi^2/\mu}} \, \left(
{\mu\over 2\pi}\right)^{12}\, ,
\ee
and, using Poisson resummation formula,
\be \label{e7}
\sum_{\vec p\in \Lambda_{E_8}} 
e^{-\mu \vec 
p^2} = \left({\pi\over \mu}\right)^4\, 
\sum_{\vec q\in \Lambda_{E_8}} e^{-\pi^2\vec q^2/\mu}
\simeq \left({\pi\over \mu}\right)^{4}\, .
\ee
Here we have used the fact that the $E_8$ lattice is self-dual.
Thus we get, for small $\mu$
\be \label{e8}
e^{\FF(\mu)} \simeq {1\over 2}\,  
\left({\mu\over 2\pi}\right)^{8} 
\, e^{{4\pi^2/\mu}} \, ,
\ee
and hence
\be \label{e9}
\FF(\mu) \simeq {4\pi^2\over \mu} + 8 \ln {\mu\over 2\pi}
+\ln{1\over 2}\, .
\ee

Before we turn to the more general case, let us try to estimate the error
in \refb{e9}. The first source of error appears in \refb{e3} where we have 
represented the symmetry requirement of the states under the $Z_2$
orbifold group by a factor of ${1\over 2}$ in
$e^\FF$. A more careful analysis described in appendix \ref{sa} shows that 
the error in $\FF$ due to this
approximation involves powers of $e^{-\pi^2/\mu}$. The second source of
error is in the small $\mu$ approximation of $\eta(e^{-\mu})$ used
in \refb{e6}. The fractional error in 
this formula is of order $e^{-4\pi^2/\mu}$. 
Finally the approximation used in \refb{e7} 
also introduces a fractional error
involving powers of $e^{-\pi^2/\mu}$. 
Thus we conclude that the net error in
eq.\refb{e9} for $\FF(\mu)$ is non-perturbative in the 
small $\mu$ approximation.

The above analysis can be easily generalized to a class of
other CHL compactifications.
We begin with heterotic string theory compactified on $T^4\times \wt S^1
\times S^1$, and tune the moduli associated with $T^4$ compactification such
that the twentyfour dimensional Narain lattice\cite{narain,nsw}
$\Lambda_{20,4}$
associated with heterotic compactification on $T^4$ has a $Z_m$
symmetry. We now
mod out the theory by a $Z_m$ symmetry group generated by
a shift $h$ of order $m$ along $\wt S^1$,
accompanied by the generator $g$ of the $Z_m$ automorphism group
of $\Lambda_{20,4}$. In order that the
final theory has $N=4$ world-sheet supersymmetry, the $Z_m$
automorphism should act trivially on the right-moving U(1) currents
of the world-sheet. However it could have non-trivial action on the
left-moving world-sheet currents, and as a result modding out by this
symmetry projects out certain number (say $k$) of the U(1) gauge fields
belonging to the Cartan subalgebra of the gauge group.
In the resulting quotient theory the rank of the gauge group associated
with the left-moving sector of the world-sheet theory is 
reduced to $(22-k)$.
If we now consider an elementary string wound $w$ times along $S^1$ and 
carrying
$n$ units of momentum along $S^1$, then the computation of 
the degeneracy $d_N$ ($N=nw$) and the partition function
$e^{\FF(\mu)}$ associated with these states involves a sum over the oscillator
levels $N_L$ as well as a sum over the $k$-dimensional momentum lattice 
$\Lambda$ whose vectors do not
couple to any massless gauge field of the resulting theory. 
This gives
\be \label{ednnew}
d_N \simeq {1\over m} \, \sum_{N_L} \, \sum_{\vec s\in \Lambda}
d_{N_L}^{(0)} \, \delta_{N_L+\vec s^2/2 -N, 1}\, ,
\ee
and
\be \label{efnew}
e^{\FF(\mu)} \simeq {1\over m} \, \sum_{N_L} d_{N_L}^{(0)} 
e^{-\mu (N_L-1)} \, \sum_{\vec s\in \Lambda} \, e^{-\mu\vec s^2/2}\, .
\ee
As discussed in appendix \ref{sa}, the factor
of $1/m$ approximately accounts for the fact that we need to count only 
those states
which are invariant under the orbifold group, and the error involved in 
this approximation involves powers of $e^{-\pi^2/\mu}$. The sum over $N_L$ 
can 
be performed using \refb{e4}, whereas the sum over $\vec s$ can be done
using Poisson resummation formula:
\be \label{epois}
\sum_{\vec s\in \Lambda} e^{-\mu \vec s^2/2} = {1\over V}
\, \left(
{2\pi\over \mu}\right)^{k/2} 
\sum_{\vec r \in \wt\Lambda} \, e^{-2\pi^2 \vec r^2/\mu} \simeq
{1\over V}\, \left(
{2\pi\over \mu}\right)^{k/2}\, ,
\ee
where $V$ denotes the volume of the unit cell in the lattice $\Lambda$
and $\wt\Lambda$ is the lattice dual to $\Lambda$.
Thus
the final expression for
$\FF(\mu)$ is given by
\be \label{e10}
\FF(\mu) \simeq {4\pi^2\over \mu} + {1\over 2} \, 
(24-k) \ln {\mu\over 2\pi} + \ln {16\over Vm}\, .
\ee
The errors in this equation come from errors in eqs.\refb{e6}, 
\refb{efnew}
and \refb{epois}. Each of these errors involves
powers of $e^{-\pi^2/\mu}$. Thus
as in the first example, for small $\mu$ the corrections to \refb{e10}
involve powers of $e^{-\pi^2/\mu}$.
 
Given $\FF(\mu)$, we define the statistical entropy $\wt S_{stat}$ 
through \refb{e11}, \refb{e12}. This gives:
\be \label{e10a}
\wt S_{stat}(N) \simeq \mu \, N +
{4\pi^2\over \mu} + {1\over 2} \, 
(24-k) \ln {\mu\over 2\pi} + \ln {16\over Vm}\, ,
\ee
where $\mu$ is the solution of the equation
\be \label{e10b}
-{4\pi^2\over \mu^2} +{24-k\over 2\mu} +N \simeq 0 \, ,
\ee
and $N=nw$.
In the limit of large $N$, the $\mu$ obtained by solving \refb{e10b} 
is
given by
\be \label{e13}
\mu \simeq {2\pi \over \sqrt{N}} 
\left(1+\OO\left({1\over \sqrt N}\right)\right) \, .
\ee
Thus for large $N$, $\mu$ is small. This justifies the small $\mu$
approximation used in arriving at \refb{e10}.
Since the error in $\FF(\mu)$ involves powers of $e^{-\pi^2/\mu}$, the
error in $\wt S_{stat}$ computed from \refb{e10a}, \refb{e10b} will
involve powers of $e^{-\pi\sqrt{N}}$.

We conclude this section by noting that $\wt S_{stat}$ computed from
\refb{e10a}, \refb{e10b} is of the form
\be \label{e14}
\wt S_{stat} = 4\pi\sqrt{N} - {24-k\over 2} \, \ln \sqrt{N} + \OO(1) \, .
\ee
Although eq.\refb{e14} gives more explicit expression for
$\wt S_{stat}$, this
equation has corrections involving inverse powers of $\sqrt N$. Thus
the comparison with the black hole entropy will be made with the formul\ae\
\refb{e10a}, \refb{e10b} for $\wt S_{stat}$ which
are correct up to error terms involving
powers of $e^{-\pi\sqrt N}$.

\sectiono{Analysis of Black Hole Entropy and Comparison with the 
Statistical Entropy} \label{sbh}

We shall now turn to the analysis of the entropy $S_{BH}$
of the BPS black hole
carrying the
same charge and mass as an elementary string state described above.
The entropy of such a black hole vanishes in the supergravity 
approximation\cite{9504147}. However the curvature associated 
with the string
metric becomes large near the horizon, showing that we must
take the higher derivative terms into account for computing
the entropy of such a black hole.
In contrast the string coupling near the horizon is small
for large
$n$ and $w$ and hence to leading order we can ignore the string loop
corrections\cite{9504147}. There is a general symmetry argument that shows 
that at the tree level in heterotic string theory the modified entropy 
must have the form $a\sqrt{nw}$ for some numerical constant 
$a$\cite{9504147,9712150,0411255}. However the value of the constant $a$ 
is not determined by this argument ($a$ could be zero for example). If 
$a=4\pi$, the black hole entropy would agree with the leading term in the 
expression \refb{e14} for $\wt S_{stat}$. Following 
the formalism developed in
refs.\cite{9812082,9904005,9906094,9910179,0007195,0009234}, 
ref.\cite{0409148} analyzed the effect of
a special class of higher
derivative terms in the tree level effective action of heterotic 
string theory which come from supersymmetric completion of the
term\cite{9602060,9603191}
\be \label{ebh0}
{1\over 16\pi} 
\int d^4 x \, \sqrt{\det g}\, \left(S\, R^-_{\mu\nu\rho\sigma} 
\, R^{-\mu\nu\rho\sigma} + \bar S \, R^+_{\mu\nu\rho\sigma} 
\, R^{+\mu\nu\rho\sigma} \right)\, ,
\ee
where $g_{\mu\nu}$, $R^\pm_{\mu\nu\rho\sigma}$ and $S$ 
denote respectively 
the canonical
metric, the self-dual and anti-self-dual components of the
Riemann tensor and the complex scalar field whose real and
imaginary parts are given by the exponential of the dilaton field 
and the axion field respectively. After taking into account the
modification of the equations of motion and supersymmetry transformation
laws due to these additional terms, the modified
black hole entropy is given by
the expression\cite{0409148,0410076,0411255,0411272}:
\be \label{eb1}
S_{BH} = {\pi N \over S_0} + 4\pi \, S_0\, , \qquad 
N \equiv nw\, ,
\ee
where $S_0$, defined as the value of the field $S$ at the horizon,
is given by the solution of the equation\footnote{Note that
the left hand side of \refb{eb2} is equal to the derivative of the 
right hand side of \refb{eb1} with respect to $S_0$. This feature
survives even after including the correction due to 
holomorphic anomaly\cite{0411255,0412287}.}
\be \label{eb2}
-{\pi\, N\over S_0^2} + 4 \pi =0\, .
\ee
This gives $S_{BH}=4\pi\sqrt{N}$. This agrees with the leading term
in the expression \refb{e14} for $\wt S_{stat}$\cite{0409148}.

Ref.\cite{0409148} checked this agreement for heterotic string 
compactification on a torus. However once this has been checked
for  torus compactification, similar agreement for other heterotic
string compactifications is
automatic due to an argument in \cite{9712150} where it was shown that at tree
level in heterotic string theory the part of the effective action relevant
for computing the entropy of these states is identical in all heterotic string
compactification with $N=4$ or $N=2$ supersymmetry. Thus the leading 
contribution
to $S_{BH}$ will be given by $4\pi\sqrt{nw}$ for all heterotic string
compactifications. This clearly agrees with the leading term in 
the expression \refb{e14} for $\wt S_{stat}$.

We now turn to the non-leading corrections to the entropy. 
For this we need to go beyond the tree level effective action of
the heterotic string theory.
A special class of
such corrections come from a term in the action of the form:
\be \label{eb3}
-{K\over 128\pi^2} \, \int d^4 x \, \sqrt{\det g}\, \ln (S+\bar S) 
\, R_{\mu\nu\rho\sigma} \, R^{\mu\nu\rho\sigma}\, ,
\ee
that arises from the so called holomorphic anomaly\cite{9302103,9309140}.
Here $K$ is a 
constant that will be determined later. (For toroidal
compactification $K=24$\cite{9610237}.)
In order to carry out a systematic analysis of the effect of this term on the
expression for the black hole entropy, we need to 
\begin{itemize}
\item supersymmetrize this term,
\item study how these additional terms in the action
modify the expression for
the black hole entropy in terms of various fields near the horizon,
\item study how the various field configurations near the horizon are
modified by these extra terms in the equation of motion, 
\item and finally evaluate the modified expression for 
the black hole entropy for the
modified near horizon field configuration.
\end{itemize}
This however has not so far been carried out explicitly. 
In order to appreciate the reason for this difficulty, one needs to
know the difference in the origin of the terms \refb{ebh0} and \refb{eb3}.
In fact both terms originate from a term of the form:\footnote{In the
convention of ref.\cite{9906094} this corresponds to choosing
$ F^{(1)} = -{i\pi\over 4} \, \left( g(S) - {K\over 64\pi^2} \ln(S+\bar S)
\right) \, \Upsilon. $
For toroidal compactification $g(S) = -{3\over 4\pi^2} \ln\eta
(e^{-2\pi S})$ 
and $K=24$.}
\be \label{eb6}
\int d^4 x \, \sqrt{\det g} \left[
\phi(S, \bar S) \, R^-_{\mu\nu\rho\sigma} 
R^{-\mu\nu\rho\sigma} + c.c.\right] \, ,
\ee
where 
\be \label{eb7a}
\phi(S,\bar S)=g(S) -{K\over 128\pi^2} \, \ln(S+\bar S)
\ee 
is the sum of a piece $g(S)$ that is holomorphic
in $S$ and a piece proportional to $\ln(S+\bar S)$
that is a function of both $S$ 
and $\bar S$.
For large $S$,
\be \label{e7b}
g(S) \simeq {S\over 16\pi}\, ,
\ee
so as to reproduce \refb{ebh0}. Hence 
\be \label{eb7}
\phi(S, \bar S)
\simeq {1\over 16\pi} \, \left( S -{K\over 8\pi} \ln(S+\bar S)
\right)\, .
\ee
A detailed analysis of the function $g(S)$ can be found in
appendix \ref{sb} where it has been shown that corrections to \refb{e7b} 
are of order $e^{-2\pi S}$. The contribution
\refb{ebh0} comes from the $g(S)\simeq S/16\pi$ term in $\phi(S,\bar S)$.
Being holomorphic in $S$,
this part is easy to supersymmetrize, and was used in arriving at expressions
\refb{eb1}, \refb{eb2} for $S_{BH}$. On the other hand \refb{eb3} arises
from the part of $\phi(S,\bar S)$ proportional to $\ln(S+\bar S)$ which cannot
be regarded as a holomorphic function. Supersymmetrization
of this term has not been carried out completely.
Nevertheless,
using various consistency requirements, \cite{9906094} guessed that 
supersymmetric completion of the term \refb{eb6} modifies
equations \refb{eb1} and \refb{eb2} to\footnote{The analysis of
\cite{9906094} was done in the context of toroidal compactification
of heterotic string theory. We are using a generalization of
this result.}
\be \label{eb4aa}
S_{BH} = {\pi N \over S_0} + 
64\pi^2 \, g(S_0) - {K\over 2} \, \ln (2S_0)\, ,
\ee
and
\be \label{eb5aa}
-{\pi\, N\over S_0^2} + 64 \pi^2 g'(S_0) -{K\over 2S_0} \simeq 0\, .
\ee
For large $N$, $S_0$ computed from \refb{eb5aa} is of order $\sqrt{N}$
and hence we can use the large $S_0$ approximation \refb{e7b} for $g(S_0)$.
This gives
\be \label{eb4}
S_{BH} \simeq {\pi N \over S_0} + 
4\pi \, S_0 - {K\over 2} \, \ln (2S_0)\, ,
\ee
and
\be \label{eb5}
-{\pi\, N\over S_0^2} + 4 \pi -{K\over 2S_0} \simeq 0\, .
\ee

{}In order to complete the computation of $S_{BH}$ we need to calculate the
constant $K$.\footnote{This calculation has been carried out earlier
in \cite{9708062} using direct analysis of one loop amplitudes in type II
string theory.} 
Fortunately there is a simple expression for $K$ by virtue of
the fact that it appears as the coefficient of the non-holomorphic piece of
$\phi(S,\bar S)$ and hence is directly related to the holomorphic anomaly
$\p_S\p_{\bar S} \phi(S,\bar S)$\cite{9302103,9309140}. 
This computation is carried out by mapping the heterotic string theory to
the dual type IIA description. As is well known, heterotic string theory
on $T^4$ is dual to type IIA string theory on 
K3\cite{9410167,9501030,9503124,9504027,9504047}.
Under this duality the Narain lattice $\Lambda_{20,4}$ of the heterotic 
string theory
gets mapped to the lattice of integer homology cycles of $K3$, and
the components of the
gauge fields take value in the real cohomology group
of $K3$\cite{9503124}. Also the generator $g$ of the $Z_m$ symmetry
of $\Lambda_{20,4}$, that was used in section \ref{sstat}
for the construction of the CHL model, gets mapped to an order $m$ symmetry
generator $\wt g$ of the conformal field theory describing type IIA string
theory on $K3$\cite{9508154} with specific action 
on the elements of the homology and the cohomology group of $K3$.
Since $g$ preserves $(24-k)$ of the 24 directions of the Narain lattice
associated with heterotic string compactification on $T^4$,
$\wt g$ will preserve $(24-k)$ of the 24 basis vectors of the real
cohomology group of $K3$. 
Now
compactifying both sides on $\wt S^1\times S^1$ we get a duality between
heterotic string theory on 
$T^4\times \wt S^1\times S^1$ and type IIA on $K3\times 
\wt S^1\times S^1$. 
Let us denote by $h$ the generator of the
order $m$ shift along $\wt S^1$. Then the CHL model, obtained by
modding out heterotic string theory on $T^4\times \wt S^1\times S^1$
by the $Z_m$ symmetry group generated by 
$h\cdot g$, is dual to
type IIA string theory on $K3\times
\wt S^1\times S^1$, modded out by the $Z_m$ group generated by
$h\cdot \wt g$\cite{9507027,9507050,9508144,9508154}.
We shall denote
by $\CC$ the conformal field theory associated with the six compact directions
of the type IIA string theory after taking this quotient. 

It is well
known that the dilaton-axion field $S$ of the heterotic string theory
gets mapped to the complexified Kahler modulus of the two torus 
$\wt S^1\times S^1$ on the type IIA side\cite{9503124}. 
Thus computation of
$\p_S\p_{\bar S}\phi(S,\bar S)$ requires computing the derivative 
of $\phi$ with
respect to the Kahler modulus of $\wt S^1\times S^1$ and its complex conjugate
in the type IIA description.
The detailed procedure for this computation can be found in 
\cite{9302103,9309140}; 
here we just summarize
the relevant result of these papers which lead to the value of $K$.
Let
us denote by $\psi^4$, $\psi^5$ the right-handed world-sheet fermions,
and by $\bar\psi^4$, $\bar \psi^5$ the left-handed world-sheet fermions 
associated with the directions along the circles $S^1$ and $\wt S^1$ in the
type IIA theory. We define:
\be \label{ec1}
\psi^\pm = {1\over \sqrt 2} (\psi^4 \pm i \psi^5), \qquad
\bar\psi^\pm = {1\over \sqrt 2} (\bar\psi^4 \pm i \bar\psi^5)\, .
\ee
In the Ramond-Ramond (RR) sector
$\psi^\pm$ as well as $\bar\psi^\pm$ have zero modes. We denote them by
$\psi_0^\pm$ and $\bar\psi_0^\pm$ respectively. 
They satisfy the usual anti-commutation relations
\be \label{ec1a}
\{\psi_0^+, \psi_0^-\} = 1, \qquad \{\bar\psi_0^+, \bar\psi_0^-\} = 1\, ,
\ee
with all other anti-commutators being zero.
If we now define
\be \label{ec2}
C = \psi_0^+ \bar \psi_0^-, \qquad \bar C = \psi_0^- \bar \psi_0^+\, ,
\ee
then in the subspace
of RR sector ground states they represent
the action of the
operators $\psi^+\bar \psi^-$ and $\psi^-\bar\psi^+$ which appear in the
vertex operator of the Kahler class of $\wt S^1\times S^1$ and its complex 
conjugate, -- the fields $S$, $\bar S$
with respect to which we want to take derivatives of
$\phi(S,\bar S)$. 
In terms of these operators the coefficient $K$ is 
given by\cite{9302103}
\be \label{ec3}
K = -Tr_{RR} \left[ (-1)^{F_L+F_R} \, C \, \bar C \right]\, ,
\ee
where the trace is to be
taken over the RR sector ground states (with $L_0=\bar 
L_0=0$) of the conformal field theory $\CC$, and $F_L$ and $F_R$ denote
the world-sheet fermion numbers in the left and the right-moving sector of
this conformal field theory. In arriving at \refb{ec3} we have used
the fact that $Tr\left( (-1)^{F_L+F_R}\right)$ vanishes in the conformal
field theory $\CC$, since the action of the fermion zero modes
$\psi_0^\pm$ pairs states with equal and opposite $(-1)^{F_L+F_R}$ 
eigenvalues.

The states of the conformal field theory $\CC$ include both untwisted sector
states and twisted sector states. Of these the twisted sector states 
necessarily carry fractional 
unit of winding along $\wt S^1$. Hence the twisted RR states
always have strictly positive $L_0$, $\bar L_0$ 
eigenvalues and cannot contribute
to \refb{ec3}. 
The untwisted sector
states are states associated with the original CFT with target
space $K3\times 
\wt S^1\times S^1$ which are invariant under the $Z_m$ 
symmetry generated by $h\cdot \tilde g$. 
These can be divided into two classes: those which are invariant
separately under $h$ and $\wt g$, and those which pick up 
equal and opposite non-trivial
phases under the action of $h$ and $\wt g$.
The latter class, being not invariant under $h$,
carries non-zero momentum along $\wt S^1$, and hence
the RR sector states in this class have strictly positive 
$L_0$ and $\bar L_0$ eigenvalues.
Thus they cannot contribute to the trace in \refb{ec3}, and we are left with
states which are invariant separately under $h$ and $\wt g$. 
Since the 
operators $C$ and $\bar C$ appearing in \refb{ec3} act on the Hilbert space
of the CFT with target space $\wt S^1\times S^1$, the contribution to $K$
from these states may be factorized as
\be \label{ec4}
K = -Tr_{RR}^{K3;inv}\left[(-1)^{F_L+F_R}\right] \, 
Tr_{RR}^{\wt S^1\times S^1} 
\left[(-1)^{F_L+F_R} C \bar C\right] \, ,
\ee
where in $Tr_{RR}^{K3;inv}$
the trace is now taken over the $\wt g$ invariant RR sector 
ground states of the
conformal field theory with target space $K3$, and in
$Tr_{RR}^{\wt S^1\times S^1}$ the trace is to be taken over the
RR sector ground states of the CFT with target space $\wt S^1\times S^1$.
For the later CFT the requirement of vanishing $L_0$ and $\bar L_0$
forces the states to carry vanishing momenta along $S^1$ and $\wt S^1$ and
hence they are automatically invariant under $h$.

There is a one to one map between the vector space of
RR sector ground states in the CFT
associated with K3 and the
real cohomology group of $K3$. Under this map $(F_L+F_R)$ get mapped to
the degree of the cohomology element. Since K3 has non-trivial
cohomology of even degree only,
$Tr\left[(-1)^{F_L+F_R}\right]$ for K3 is equal to the dimension of the 
cohomology group of K3 which is 24. Here however we are interested in the
RR sector ground states which are even under $\wt g$, and hence we should
count the dimension of the cohomology group of K3 which is invariant under 
$\wt g$. Since this number is equal to $(24-k)$, we have
\be \label{ec5}
Tr_{RR}^{K3;even}\left[(-1)^{F_L+F_R}\right] = 24-k\, .
\ee
In order to calculate the second factor appearing in \refb{ec4}, we note that
the RR sector ground states 
associated with $\wt S^1\times S^1$ consist of four
states. Defining the vacuum state $|0\ra$ to
be annihilated by $\psi_0^-$ and
$\bar\psi_0^-$, and have $F_L=F_R=-{1\over 2}$, the states are
\be \label{ec6}
|0\ra, \quad \psi_0^+|0\ra, \quad \bar\psi_0^+|0\ra, \quad \psi_0^+\bar
\psi_0^+|0\ra \, , 
\ee
with $(F_L,F_R)$ values $\left(-{1\over 2}, -{1\over 2}\right)$, 
$\left(-{1\over 2}, {1\over 2}\right)$, $\left({1\over 2}, -{1\over 2}\right)$, 
$\left({1\over 2}, {1\over 2}\right)$ respectively. {}From the structure
of $C$ and $\bar C$ defined in \refb{ec2} it is clear that only the
state $\psi_0^+|0\ra$ will contribute to the trace appearing in
the second factor of \refb{ec4}. Since $C\bar C\psi_0^+|0\ra=-\psi_0^+|0\ra$,
we get
\be \label{ec7}
Tr_{RR}^{\wt S^1\times S^1} 
\left[(-1)^{F_L+F_R} C \bar C\right] =-1\, .
\ee
Substituting \refb{ec5} and \refb{ec7} into \refb{ec4}  we get
\be \label{ec8}
K = (24-k)\, .
\ee

Eqs.\refb{eb4}, \refb{eb5} now give
\be \label{eb4a}
S_{BH} \simeq {\pi N \over S_0} + 4\pi \, S_0 - {1\over 2}
\, (24-k) \, 
\ln (2S_0)\, ,
\ee
and
\be \label{eb5a}
-{\pi\, N\over S_0^2} + 4 \pi -{24-k\over 2S_0} \simeq 0\, .
\ee
These agree with eqs.\refb{e10a}, \refb{e10b} under the identification
\be \label{ec9}
\wt S_{stat} = S_{BH}, \qquad \mu = {\pi\over S_0}\, .
\ee
Thus we see that in this approximation the entropy $S_{BH}$
of the black hole
agrees with the statistical entropy $\wt S_{stat}$ calculated following
the procedure given in section \ref{srev} up to an overall constant.

Earlier we had estimated the error in $\wt S_{stat}$ calculated from
\refb{e10a}, \refb{e10b} to be
nonperturbative in $1/\sqrt{N}$. We shall now try to carry out a
similar estimate of the error involved in eqs.\refb{eb4a}, \refb{eb5a} so
that we can determine up to what level the agreement between $S_{BH}$ and
$\wt S_{stat}$ holds.
First of all we should remember that there is an
uncertainty involved in the
original formulae \refb{eb4aa}, \refb{eb5aa} since they have not been derived
from first principles. 
In particular \cite{9906094} used an argument based on duality symmetry to
derive the effect of the holomorphic anomaly terms, and this does not
fix the additive constant on the right hand side
of \refb{eb4aa}. Thus there is an ambiguity in the overall additive
constant in the
expression for $S_{BH}$, and hence we cannot hope to compare the additive
constants in $\wt S_{stat}$ and $S_{BH}$.
Assuming that the formul\ae\ \refb{eb4aa}, \refb{eb5aa} are correct up to this
additive constant, we see
that the error in the determination of $S_{BH}$ lies essentially in the
error in the determination of the function $g(S)$. As reviewed in 
appendix \ref{sb}, 
we can determine the form of the corrections to $g(S)$ by the requirement 
of S-duality invariance of the
theory,\footnote{As will be discussed in appendix \ref{sb},
the S-duality group for these models is usually
a subgroup of SL(2,Z), and 
depends on the specific
model we are analysing\cite{9507050}.}
and typically corrections
to \refb{e7b} involve powers of $e^{-2\pi S}$. Since $S_0$ obtained by 
solving eq.\refb{eb5a} is of order
$\sqrt{N}$, we see that the corrections to $S_{BH}$ will involve powers of
$e^{-\pi\sqrt{N}}$. Thus for large $N$
the agreement between $S_{BH}$ and $\wt S_{stat}$ 
holds up to an undetermined additive
constant, and terms which are non-perturbative in $1/\sqrt{N}$.
In particular if we express $\wt S_{stat}$ and $S_{BH}$ in power series
expansion in $1/\sqrt{N}$ by solving eqs.\refb{e10a}, \refb{e10b} 
and \refb{eb4a}, \refb{eb5a}
respectively, then the results agree to all orders in $1/N$ including
terms proportional to $\ln N$. This in turn implies similar agreement
between $\ln d_N$ and $\ln d_N^{BH}$ defined in section \ref{srev}.

\sectiono{Discussion} \label{sdiss}

Given the agreement between the statistical entropy $\wt S_{stat}$ and 
the black hole entropy $S_{BH}$ up to non-perturbative terms, one might 
wonder if this correspondence also holds after we include non-perturbative 
terms. Unfortunately however even for toroidally compactified heterotic 
string theory $\wt S_{stat}$ and $S_{BH}$ differ once we include
non-perturbative corrections\cite{0411255,0412287}, and hence we
expect that such disagreement will also be present
for CHL compactifications. One could contemplate several reasons for
this discrepancy:
\begin{enumerate}
\item First of all we should remember that the formul\ae\
\refb{eb4aa},
\refb{eb5aa}
for the black
hole entropy in the presence of non-holomorphic terms, as given in
\cite{9906094}, have not been derived from first principles. Thus
there could be further corrections to these formul\ae\ which could
modify the expression for $S_{BH}$.
\item It could also be that the proposal \refb{e2a} - \refb{eprop}
for relating the black hole
entropy to the degeneracy of elementary string states
is not 
complete; and that the formul\ae\ needs to be
modified once non-perturbative effects are taken into account.
\item Besides supersymmetric
completion of the curvature squared terms, the effective field
theory contains infinite number of other higher derivative terms
which are in principle equally important, and
at present there is no understanding as to why these terms do not
affect the expression for the entropy. It could happen that while
these terms do not affect the perturbative corrections, their
contribution becomes significant at the non-perturbative level.
\item Finally there is always a possibility that the relation between
the black hole entropy and statistical entropy exists only as a
power series expansion in inverse powers of various charges. In this
case we do not expect any relation between the non-perturbative
terms in the expressions for $S_{BH}$ and $\wt S_{stat}$. 
\end{enumerate}

At present we do not know which (if any) of these possibilities
is correct. This issue clearly needs further investigation.
We note however that if the fourth proposal is correct, namely the 
agreement between the black hole entropy and the statistical entropy holds 
only as a power series expansion in inverse powers of the charges, then 
the proposal \refb{eprop} relating the black hole and the statistical 
entropy can be extended to more general models and more general states. 
The essential point is that in our analysis we have been restricted to 
compactifications of the form $K_5\times S^1$ and to states carrying 
momentum and winding along $S^1$ in order to ensure that the degeneracy of 
the states depends only on the 
combination $N=nw$. If
we consider more general $N=4$ supersymmetric compactification 
({\it e.g.} where the orbifold group is $Z_m\times Z_{m'}$
and acts on both circles instead of just 
one circle\cite{9508154}) and/or more general 
states 
carrying arbitrary gauge charges $(\vec P_L, \vec P_R)$ associated
with gauge fields arising out of the left and the right sectors of the
world-sheet, then the role of the T-duality invariant combination $nw$ is 
played by
$N\equiv {1\over 2} (\vec P_R^2 - 
\vec P_L^2)$. However in this case the degeneracy $d(\vec P_L, \vec P_R)$
of such states could depend on $\vec P_L$ and $\vec P_R$ separately
instead of being a function of the combination $N$ only. To see how
such dependence can arise, we can
consider the class of models described in section \ref{sstat} and consider
a state that carries $\tilde n$ units of momentum along the circle $\wt S^1$ 
besides the charge
quantum numbers $n,w$ associated with the circle $S^1$. 
For such states we still have $N=nw$. However
in this case the
part of the wave-function associated with $\wt S^1$
picks up a phase $e^{2\pi i \tilde n/m}$ 
under the $Z_m$ shift along $\wt S^1$,
and we must compensate for it by introducing a factor of 
$e^{2\pi i \tilde n l/m}$
multiplying $g^l$ in the projection operator \refb{epr1} in order to 
ensure that the complete state is invariant under the $Z_m$ 
transformation. This introduces
a specific dependence of the partition function and hence of
the degeneracy of states on $\tilde n$. If in addition the state carries
some gauge charges associated with the lattice $\Lambda_{20,4}$, then
the sum over
momentum $\vec s$ in \refb{ednnew}, \refb{efnew} 
might run over a shifted lattice,
which is equivalent to replacing the $\exp(-\mu\vec s^2/2)$ factor
in \refb{efnew} by $\exp[-\mu (\vec s + \vec K)^2/2]$ for some fixed vector
$\vec K$ that depends on the component of
$(\vec P_L, \vec P_R)$ along $\Lambda_{20,4}$. However by following the
analysis given in section \ref{sstat} and
appendix \ref{sa} one can see that the
dependence on $(\vec P_L, \vec P_R)$ introduced by either of these effects
is exponentially suppressed, and
hence if we are interested in $\ln d(\vec P_L, \vec P_R)$ as a power series
expansion in inverse powers of charges, the result depends only on the
combination $N=(\vec P_R^2-\vec P_L^2)/2$. Similar analysis can also
be carried out for the $Z_m\times Z_{m'}$ orbifold models.
This allows us to define
$\wt S_{stat}(N)$ through eqs.\refb{e2a}-\refb{e12} within this approximation.
On the other hand from the results of \cite{9906094} it 
also follows that the black hole entropy will also be a function of the 
combination $N={1\over 2} (\vec P_R^2 -
\vec P_L^2)$. An analysis similar to the one described in this paper
can then be used to show that
the correspondence \refb{eprop} between the statistical 
entropy and 
the black hole entropy continues to hold for these more general class
of states and/or models.

{\bf Acknowledgement}: I would like to thank P.~Aspinwall, A.~Dabholkar,
R.~Gopakumar, D.~Jatkar
and D.~Surya Ramana for valuable discussions.

\appendix

\sectiono{Estimating Error in the Computation
of Statistical Entropy} \label{sa}

In this appendix we shall estimate
corrections to eq.\refb{efnew} arising
out of the fact that we needed to sum over $g$ 
invariant states with $g$ being the generator of a $Z_m$ group; but instead
we summed over all states and divided the result by a factor of $m$. The
correct expression would involve inserting in the sum over states a 
projection operator 
\be \label{epr1}
{1\over m} \sum_{l=0}^{m-1} \, g^l\, 
\ee
that projects onto $g$ invariant states. The $l=0$ term in the
above expression reproduces the right hand side of \refb{efnew}. Thus
we need to estimate the contribution due to the $l\ne 0$ terms.

$g$ acts non-trivially on the left-moving
oscillators as well as the internal momentum carried by the state. We can 
choose appropriate linear combination of the left moving world-sheet
scalar fields so that the 
annihilation operator $a_{\alpha n}$ 
($1\le \alpha\le 24$, $1\le n<\infty$)
associated with the $\alpha$-th
scalar field picks up a $Z_m$ phase $e^{2\pi i k_\alpha / m}$ 
($0\le k_\alpha \le (m-1)$) under the $g$ transformation.
In that case the action
of $g$ on the oscillators is represented by the operator
\be \label{epr2}
\exp\left( {2\pi i \over m} \sum_{\alpha=1}^{24} 
\sum_{n=1}^\infty k_\alpha \, a^\dagger_{\alpha n} a_{\alpha n}
\, \right)\, .
\ee
Let $\wh g$ denote the action of $g$ on the internal momentum. Then
the $l>0$ terms in the sum in \refb{epr1} have the form
\be \label{epr4}
{1\over m} \sum_{l=1}^{m-1}
\exp\left( {2\pi i l \over m} \sum_{\alpha=1}^{24}
\sum_{n=1}^\infty  k_\alpha \, a^{\dagger}_{\alpha n} 
a_{\alpha n} \right)\, \wh g^l\, .
\ee
Inserting this into the trace over BPS states weighted by 
$e^{-\mu(N_L-1+\vec s^2/2)}$ we see first of all that 
unless $\wh g^l \vec s= \vec s$ the lattice vector $\vec s$ does not
contribute to the trace. Thus the momentum
sum receives
contribution only from a sublattice $\Lambda_l$ of
$\Lambda$ which is invariant under $\wh g^l$.
Since $\Lambda$ is transverse to the directions in $\Lambda_{20,4}$
which are left invariant under $\wh g$, 
$\wh g$ does not preserve any direction of the lattice $\Lambda$. Thus
for $l=1$ the sublattice $\Lambda_l$ consists of the single point 
$\vec s=0$. But if $m$ is not prime, then $\wh g^l$ can be of
order $<m$ for some $l$ and $\Lambda_l$
could be non-trivial.
On the other hand, contribution to the $l$-th term in $e^{\FF(\mu)}$
from the oscillators
($a_{\alpha n}, a^{\dagger}_{\alpha n}$) is given by
\be \label{eosctr}
Tr_{\alpha,n} \left( e^{-\mu n a^{\dagger}_{\alpha n} \,
a_{\alpha n} + 2\pi i l k_\alpha a^{\dagger}_{\alpha n} \,
a_{\alpha n} / m }\right) = 
{1\over 1-e^{2\pi i l k_\alpha / m} e^{-n\mu}}
\, ,
\ee 
where $Tr_{\alpha,n}$ denotes trace over all states created by (multiple)
application of the oscillator $a^{\dagger}_{\alpha n}$ on the vacuum.
Thus the net 
additional contribution to $e^{\FF(\mu)}$ from the $l\ne 0$ terms is
given by:
\be \label{epr5}
{1\over m} \sum_{l=1}^{m-1} \prod_{\alpha=1}^{24} \prod_{n=1}^\infty
{1\over 1-e^{2\pi i l k_\alpha / m} q^n } \,
\sum_{\vec s\in \Lambda_l} e^{-\mu
\vec s^2/ 2}\, , \qquad q\equiv e^{-\mu}\, .
\ee
We need to compare this with the $l=0$ term. First of all we see that
contribution from the momentum sum, running over a sublattice of $\Lambda$,
is suppressed by a power of $\mu$ that depends on the difference in
the dimension of the original lattice $\Lambda$ and the sublattice
$\Lambda_l$. But more importantly, the contribution from the $\alpha$-th
oscillator is suppressed by a factor of
\be \label{epr6}
\prod_{n=1}^\infty
{1\over 1-e^{2\pi i l k_\alpha / m} q^n }  \bigg/
\prod_{n=1}^\infty
{1\over 1-q^n }  \, .
\ee
For small $\mu$ this ratio is suppressed by a factor of\cite{9405117}
\be \label{esupp}
\left({4\pi \over \mu} \sin{\pi k_\alpha l\over m} \right)^{1/2}
\exp\left[ -{\pi^2\over \mu} \, {k_\alpha l\over m} 
\left(1 - {k_\alpha l\over m}
\right)\right] \, ,
\ee
for $0<k_\alpha l/m < 1$.
This
shows that the correction to $\FF(\mu)$ given in \refb{efnew} involves
powers of $e^{-\pi^2/\mu}$. The argument given below eq.\refb{e13} now leads
to the conclusion that the error in $\wt S_{stat}$ computed from \refb{e10a},
\refb{e10b} involves powers of $e^{-\pi\sqrt N}$.

\sectiono{Estimating Error in the Computation
of Black Hole Entropy} \label{sb}

In this appendix we shall analyze the function $g(S)$
and estimate the error in the expression 
\refb{e7b} for this function.\footnote{For some orbifold models
similar analysis has been carried out in \cite{9708062}.}
The basic tool used in this analysis
will be S-duality invariance. 
The S-duality group of 
the CHL 
models of the type considered in this paper may be
found by identifying it as the T-duality group in the dual type
IIA description\cite{9507050}. It
is a subgroup of the SL(2,Z) group that commutes with the orbifold
group
and acts on $S$ as
\be \label{ebb1}
S\to - i {i a S + b \over icS + d}\, \qquad ad-bc = 1, \qquad a,b,c,d\in 
Z\, ,
\ee
together with some additional restriction on $a$, $b$, $c$, $d$.
For the $Z_m$ orbifold models described in section \ref{sstat}
this additional restriction takes the
form\cite{9507050}:\footnote{I wish to thank P.~Aspinwall for
discussion on this point.}
\be \label{eadd2}
\pmatrix{a & b\cr c & d} \pmatrix{1\cr 0} = \pmatrix{\hbox{1 mod $m$}
\cr \hbox{0 mod $m$}}, \qquad \i.e. \quad a=\hbox{1 mod $m$},
\quad  c=\hbox{0 mod $m$}\, .
\ee
The resulting group is known as 
$\Gamma_1(m)$. It is a subgroup of $\Gamma_0(m)$ 
containing $SL(2,\ZZZ)$ matrices of the form 
\refb{eadd2} with the
condition $a=1$ mod $m$ relaxed\cite{schoen}. It
was shown in \cite{9906094}
that in order that the supergravity effective action is
invariant under S-duality after addition of the term
\refb{eb6} to the action, the combination
\be \label{ebb2}
h(S,\bar S) \equiv \p_S\left(g(S) -{K\over 64\pi^2} \, \ln(S+\bar 
S)\right)
\ee
must transform under a duality 
transformation as:
\be \label{ebb3}
h\left(- i {i a S + b \over icS + d}, i {-i a \bar S + b \over -ic\bar S + 
d} \right) = (i c S + d)^2 h(S, \bar S)\, .
\ee
{}Using the known modular transformation laws of $\eta\left(e^{-2\pi 
S}\right)$ and $S+\bar S$ it then follows that the holomorphic
combination
\be \label{ebb4}
\wh h(S) \equiv h(S, \bar S) +{K\over 64\pi^2} \, \p_S \left[
\ln \eta^2\left(e^{-2\pi S}\right) + \ln(S+\bar S)\right] =
\p_S g(S) +{K\over 32\pi^2}\,  \p_S \ln \eta\left(e^{-2\pi 
S}\right)\,
\ee
transforms as a modular form of weight two:
\be \label{ebb5}
\wh h\left(- i {i a S + b \over icS + d}
\right) = (i c S + d)^2 \wh h(S)\, .
\ee
Furthermore, from \refb{e7b} and 
\refb{ebb4} it follows 
that for large $S$,
\be \label{ebb6}
\wh h(S) \simeq {1\over 16\pi} \, \left ( 1 - {K\over 24}\right)\, .
\ee

For toroidal compactification of heterotic string theory the S-duality
group is SL(2,Z) which has no modular form of weight two. Thus $\wh h(S)$
must vanish. This is consistent with eq.\refb{ebb6} since $K=24$ for
toroidal compactification. In general however $K$ given in \refb{ec8}
is less than 24, and hence $\wh h(S)$ must be non-trivial. Fortunately
$\Gamma_0(m)$ (and hence $\Gamma_1(m)\subset \Gamma_0(m)$)
for $m\ge 2$ does have non-trivial 
modular forms of weight
two\cite{schoen}. If $m$ is prime then this modular form is unique and
explicit expression 
for this modular form is given by\cite{schoen}:
\be \label{emodu}
E(iS; m) = m G_2^*(imS) - G_2^*(iS) \, ,
\ee
where
\be \label{emodu1}
G_2^*(iS) = {\pi^2\over 3} - 8\pi^2 \, \sum_{n_1,n_2\ge 1}
n_1 e^{-2\pi n_1 n_2 S} = -4\pi \p_S \ln \eta\left(e^{-2\pi S}\right)\, .
\ee
$E(iS;m)$ is normalized such that for $S\to\infty$, $E(iS;m)\to
(m-1)\pi^2/3$. Thus eq.\refb{ebb6} gives
\be \label{emodu2}
\wh h(S) ={3\over 16\pi^3 (m-1)} \, \left( 1 - {K\over 24}\right)
E(iS; m)\, .
\ee
$g(S)$ can now be determined from \refb{ebb4} up to an overall additive
constant.

If $m$ is not prime then $\wh h(S)$ is not determined uniquely by
this argument. Nevertheless
since modular forms of $\Gamma_0(m)$ have series
expansion in powers of 
$e^{-2\pi S}$, we see 
that the 
correction to \refb{ebb6} is of order $e^{-2\pi 
S}$. This in turn implies that the 
corrections to \refb{e7b} involve powers of $e^{-2\pi S}$ and 
hence are non-perturbative in $1/S$. We can now invoke the analysis
below eq.\refb{ec9} to conclude that the error in $S_{BH}$ computed from
eqs.\refb{eb4a}, \refb{eb5a} involves powers of $e^{-\pi\sqrt N}$.

\end{document}